\documentclass[9pt,technote, final]{IEEEtran}

\usepackage{bm}
\usepackage{cite}
\usepackage{tikz}
\usepackage{array}
\usepackage{float}
\usepackage{lineno}
\usepackage{comment}
\usepackage{calrsfs}
\usepackage{amssymb}
\usepackage{makecell}
\usepackage{textcomp}
\usepackage{graphicx}
\usepackage{epstopdf}
\usepackage{amsfonts}
\usepackage{calligra}
\usepackage{mathrsfs}
\usepackage{mathtools}
\usepackage{subcaption}
\usepackage{color, soul}
\usepackage{lipsum,adjustbox}
\usepackage{tabularx,colortbl}
\usepackage{letltxmacro,xparse}
\usepackage[hidelinks]{hyperref}
\usepackage[hang,flushmargin]{footmisc}

\usetikzlibrary{shapes}
\usetikzlibrary{shapes.geometric, arrows, fit}

\graphicspath{{figures/}}
\definecolor{yellow}{RGB}{255,255,0}
\definecolor{red}{RGB}{255,91,51}
\definecolor{blue}{RGB}{8,191,223}
\definecolor{green}{RGB}{54,203,30}
\definecolor{grey}{RGB}{170,170,170}
\definecolor{black}{RGB}{0,0,0}
\definecolor{orange}{RGB}{255,140,0}
\definecolor{yellow}{RGB}{225,249,27}
\tikzset{VertexStyle/.style = {shape = rectangle,fill = gray}}

\def\footnoterule{\relax%
 \kern15pt
 \hbox to \columnwidth{\hfill\vrule width 1\columnwidth height 0.6pt\hfill}
 \kern4.6pt}

\LetLtxMacro\oldhref\href
\RenewDocumentCommand{\href}{o m m}{%
 \IfValueTF{#1}
 {\oldhref[#1]{#2}{\bfseries #3}}
 {\oldhref{#2}{\bfseries #3}}%
}

\DeclareMathOperator*{\argmin}{argmin}
\DeclareMathOperator*{\argmax}{argmax}

\newcommand{\apre}{{\textit{a priori }}}

\tikzstyle{arrow} = [thick,->,>=stealth]
\tikzstyle{circ} = [circle, minimum width=1cm, minimum height=1cm, text centered, draw=black, fill=blue!30]
\tikzstyle{circ_open} = [circle, minimum width=1cm, minimum height=1cm, text centered, draw=black, fill=red!60]
\tikzstyle{decision} = [diamond, minimum width=3cm, minimum height=1cm, text centered, draw=black, fill=orange!50,font=\bfseries]
\tikzstyle{process} = [rectangle, minimum width=6.0cm, minimum height=1cm, text centered, draw=black, fill=blue!60,font=\bfseries]
\tikzstyle{startstop} = [rectangle, rounded corners, minimum width=4.5cm, minimum height=1cm,text centered, draw=black, fill=green!50,font=\bfseries]

\usepackage{acronym}

    \acrodef{WVU}{West Virginia University}
    \acrodef{MAE}{Mechanical and Aerospace Engineering}

    \acrodef{IMU}{inertial measurement unit}
    \acrodef{INS}{inertial navigation system}
    \acrodef{GPS}{global positioning system}
    \acrodef{GNSS}{global navigation satellite system}
    \acrodef{LiDAR}{light detection And ranging}
    
    \acrodef{PLL}{phase lock loop}
    \acrodef{DLL}{delay lock loop}
    \acrodef{IQ}{in-phase and quadrature}
    \acrodef{SDR}{software defined radio}
    \acrodef{RINEX}{receiver independent exchange format}
    \acrodef{RTK}{real time kinematic}
    \acrodef{PPP}{precise point positioning}
    
    \acrodef{SQM}{signal quality metric}
    \acrodefplural{SQM}[SQMs]{signal quality metrics}
    \acrodef{SS}{signal strength}
    \acrodef{EL}{elevation angle}
    \acrodef{AZ}{azimuth angle}

    \acrodef{KF}{Kalman filter}
    \acrodef{EKF}{extended Kalman filter}
    \acrodef{GMM}{Gaussian mixture model}
    \acrodef{NLLS}{nonlinear least squares}
    \acrodef{SLAM}{simultaneous localization and mapping}
    \acrodef{MAP}{maximum a posteriori}
    \acrodef{LARS}{least angle regression}
    
    \acrodef{m-estimator}{maximum likelihood estimator}
    \acrodefplural{m-estimator}[m-estimators]{maximum likelihood estimator}
    
    \acrodef{IRLS}{iteratively re-weighted least squares}
    \acrodef{DCS}{dynamic covariance scaling}
    \acrodef{MM}{max-mixtures}
    \acrodef{BCE}{batch covariance estimation}
    \acrodef{BCE-AD}{batch covariance estimation over an augmented data-space}
    \acrodef{RANSAC}{random sample consensus}
    \acrodef{RRR}{realizing, reversing, recovering}
    \acrodef{RAIM}{receiver autonomous integrity monitoring}

    \acrodef{MAP}{maximum a posteriori}
    \acrodef{MLE}{maximum likelihood estimate}

    \acrodef{RSOS}{residual-sum-of-squares}
    \acrodef{HRSOS}{horizontal residual-sum-of-squares}
    \acrodef{TRSOS}{total residual-sum-of-squares}
    
\acrodef{NN}{nearest neighbor}
\acrodef{VDP}{variational Dirichlet process}

\begin{document}


\title{Uncertainty Model Estimation in an Augmented Data Space for Robust State Estimation}

\author{Ryan~M.~Watson, Jason~N.~Gross, Clark~N.~Taylor,~and~Robert~C.~Leishman%
\thanks{R. Watson and J. Gross are with the Department of Mechanical and Aerospace Engineering, West Virginia University, 1306 Evansdale Drive, PO Box 6106, WV 26506-6106, United States. e-mail: rwatso12 (at) mix.wvu.edu. \newline}
\thanks{C. Taylor is with the Department of Electrical and Computer Engineering, Air Force Institute of Technology, 2950 Hobson Way, WPAFB, OH 45433-7765, United States. \newline}
\thanks{R. Leishman is with the Autonomy and Navigation Technology Center, Air Force Institute of Technology, 2950 Hobson Way, WPAFB, OH 45433-7765, United States.\newline}}


\maketitle
\markboth{Correspondence Submitted to IEEE Transactions on Aerospace and Electronic Systems}%
{Watson \MakeLowercase{\textit{et al.}}: Uncertainty Model Estimation in an Augmented Data Space for Robust State Estimation}

\begin{abstract}
The requirement to generate robust robotic platforms is a critical enabling step to allow such platforms to permeate safety-critical applications (i.e., the localization of autonomous platforms in urban environments). One of the primary components of such a robotic platform is the state estimation engine, which enables the platform to reason about itself and the environment based upon sensor readings. When such sensor readings are degraded traditional state estimation approaches are known to breakdown. To overcome this issue, several robust state estimation frameworks have been proposed. One such method is the batch covariance estimation (BCE) framework. The BCE approach enables robust state estimation by iteratively updating the measurement error uncertainty model through the fitting of a Gaussian mixture model (GMM) to the measurement residuals. This paper extends upon the BCE approach by arguing that the uncertainty estimation process should be augmented to include metadata (e.g., the signal strength of the associated GNSS observation). The modification of the uncertainty estimation process to an augmented data space is significant because it increases the likelihood of a unique partitioning in the measurement residual domain and thus provides the ability to more accurately characterize the measurement uncertainty model. The proposed batch covariance estimation over an augmented data-space (BCE-AD) is experimentally validated on collected data where it is shown that a significant increase in state estimation accuracy can be granted compared to previously proposed robust estimation techniques. 
\end{abstract}

\section{Introduction} \label{sec:intro}

The expected operation domain of autonomous robotic platforms is ever-increasing. Currently, these applications span the space from structured industrial applications to autonomous exploration of novel environments~\cite{mallios2016toward}. This expectation has lead to astonishingly complex robotic systems with stringent safety requirements.

One of the key components of such robotic systems is the ability to accurately infer the essential states of the system when provided with a set of information (i.e., state estimation~\cite{barfoot2017state}). When the provided information adheres to the \apre models, this problem is addressed through numerous optimization frameworks (i.e., Kalman filtering~\cite{simon2006optimal}, particle filtering~\cite{thrun2005probabilistic}, or graphical methods~\cite{dellaert2017factor}). However, many scenarios emit characteristics that do not adhere to the specified assumption (e.g., the accuracy of the \apre models cannot be guaranteed in an novel operating environment). In such situations, the solution provided by any of the mentioned estimation frameworks can be arbitrarily biased by any single unmodeled observations\footnote{This can be shown through the calculation of the asymptotic breakdown point of the $l^2$-based estimators (i.e., any estimation framework that utilizes the $l^2\text{-norm}$ exclusively to construct the objective function)~\cite{hampel1968contribution}.}.

To combat the breakdown of the $l^2\text{-based}$ estimators, research has been conducted in robust state estimation (i.e., state estimation methodologies that enable an increased breakdown point~\cite{hampel1968contribution} when compared to traditional $l^2\text{-based}$ estimation). These robust estimation frameworks can be categorized into two paradigms: data weighting techniques, and data exclusion techniques. The data weighting techniques (e.g., robust \acp{m-estimator}~\cite{huberBook}, \ac{DCS}~\cite{DCS}, and \ac{MM}~\cite{maxmix}) enable robustness by reducing the influence (i.e., increase the corresponding uncertainty) of data which does not adhere to the \apre models. The data exclusion techniques (e.g., \ac{RANSAC}~\cite{fischler1981random}, \ac{RAIM}~\cite{zhai2018fault}, and $l^1\text{-relaxation}$~\cite{carlone2014selecting}) enable robustness through selecting a subset of the data (i.e., the subset of data that adheres to the \apre models) and conducting state estimation with only the trusted subset.

All of the robust state estimation implementations described above share one key deficiency. Specifically, they all assume that the \apre uncertainty model accurately characterize the underlying system. This assumption can be problematic, for example, in adversarial environments that emit data degrading characteristics (i.e., environments where it is not feasible to accurately know \apre the measurement uncertainty model). To overcome this issue, the \ac{BCE} approach was proposed within~\cite{watsonbatch, watson2019enabling}. The \ac{BCE} approach works by iteratively updating the measurement uncertainty model between each iteration of optimization by fitting a \ac{GMM} to the measurement residuals. This approach has the benefit of learning the underlying system uncertainty model instead of assuming it from the outset. 

The work presented within this paper provides an extension to the \ac{BCE} approach. Specifically, this work removes the assumption that the measurement uncertainty model can be accurately characterized exclusively utilizing information contained within the measurement residuals. Instead, this work argues that the uncertainty estimation problem can effectively be augmented to incorporate metadata, leading to a more accurate characterization of the observation uncertainty model. As shown in Section \ref{sec:experimental_validation} the increase in positioning accuracy achieved by including metadata in the residual classification procedure can be significant.

The remainder of this paper proceeds in the following manner. In section \ref{sec:state_est}, a succinct overview of \ac{NLLS} is provided, with a specific emphasis placed on robust estimation. In Section \ref{sec:proposed_approach}, the \ac{BCE-AD} robust estimation methodology is discussed. In Section \ref{sec:experimental_validation} the proposed approach is validated with multiple kinematic \ac{GNSS} data-sets. Finally, the paper concludes with final remarks and proposed future research efforts.

\section{State Estimation} \label{sec:state_est}

In this section, we briefly review of state estimation and its robust variants. For a more thorough discussion, the reader is referred to Section II of~\cite{watson2019enabling}. To begin the discussion, the state estimation problem can generically be defined as the process of calculating a set of states $X$ that -- in some sense -- are in best agreement with the provided information $Y$. Within this work, the metric utilized to quantify agreement is the maximization of the \textit{posterior} distribution (i.e., the \ac{MAP} state estimate $\hat{X}$), as presented in Eq. \ref{eq:map_cost}. 

\begin{equation}
 \hat{X} = \argmax_X \ \operatorname{p}(X \ | \ Y)
 \label{eq:map_cost}
\end{equation}

The implementation a \ac{MAP} estimator can be achieved through the utilization of the factor graph~\cite{dellaert2017factor} formulation. The factor graph is a probabilistic graphical model framework which enables the factorization of the \textit{posterior} distribution into a product of functions that operate on a reduced domain, as shown in Eq. \ref{eq:post_factorization}

\begin{equation}
\operatorname{p}(X \ | \ Y) \propto \prod_{n=1}^{N} \psi_n(A_n,B_n),
\label{eq:post_factorization}
\end{equation}

\noindent where, $\psi_n(A_n,B_n)$ is an application specific domain reduced function (i.e., a factor in the factor graph model), which operates on $A_n \subseteq \{ X_1, X_2 \ldots, X_n\}$, and $B_n \subseteq \{ Y_1, Y_2 \ldots Y_m \}$. 

To facilitate a computationally efficient implementation, it is commonly assumed that each factor within the models adheres to a Gaussian noise assumption. When this assumption is enforced, the estimation problem presented in Eq. \ref{eq:map_cost} is reduced to finding the set of states which minimizes the squared sum of weighted residuals~\cite{dellaert2006square}, as presented in Eq. \ref{eq:nlls_cost}

\begin{equation}
 \hat{X} = \argmin_X \sum_{n=1}^{N} \lvert \lvert \ r_n(X) \ \rvert \rvert_{\Lambda_n} \quad \text{s.t.} \quad r_n(X) \triangleq y_n - h_n(X),
 \label{eq:nlls_cost}
\end{equation}

\noindent where $r_n(X)$ is an observation residual, $h_n$ is a function that maps the state estimate to the observation domain, $\Lambda_n$ is a covariance (i.e., residual weighting) matrix, and $|| \ * \ ||$ is defined as the $l^2\text{-norm}$.

\subsection{Robust State Estimation}

Based upon the discussion provided within~\cite{watson2019enabling}, we can view robust state estimation schemes as variants of \ac{IRLS}~\cite{zhang1997parameter}. The implementation of the \ac{IRLS} formulation is presented in Eq. \ref{eq:irls}, where $w_n$ is a real valued function that attempts to appropriately weight observations based upon the previous iteration's residuals.

\begin{equation}
    \hat{X} = \argmin_X \sum_{n=1}^N w_n(e_{n}) || r_n(X) ||_{\Lambda_n}  \quad \text{s.t.} \quad e_n \triangleq \lvert \lvert r_{n-1}(X) \rvert \rvert_{\Lambda_n}
    \label{eq:irls}
\end{equation}

The specific implementation of weighting functions has been extensively explored. One such implementation, commonly utilized within the robotics community, is the \ac{DCS} approach~\cite{DCS} which implements a redescending\footnote{A weighting function that approaches $0$ as the magnitude of the error approaches $\infty$ (i.e., $\lim_{e_n \to \infty} w(e_n) = 0$)\newline} \acp{m-estimator} type weighting function. Another commonly utilized framework is the \ac{MM} approach~\cite{maxmix}, which enable the utilization of a \ac{GMM} error uncertainty model through the approximation of the summation operation with the maximum operation\footnote{When a \ac{GMM} is utilized to characterize the error uncertainty model, the objective function of the corresponding estimation problem does not reduce to a \ac{NLLS} form, which increases computation complexity. To reduce this additional complexity, the maximum operation can be utilized in place of the summation operation in the \ac{GMM} (i.e., $\operatorname{p}(y_i \ | \ X) = \sum_{m=1}^{M} w_m \mathcal{N}(\mu_m, \Lambda_m) \approx \max_i w_i \mathcal{N}(\mu_i, \Lambda_i))$.}. Finally, the \ac{MM} approach was extended within~\cite{watsonbatch,watson2019enabling,pfeifer2018expectation} to allow for non-parametrically learned uncertainty models (i.e., the uncertainty model is not assumed to be static) based upon the measurement residuals in the \ac{BCE} approach.
\section{Proposed Approach} \label{sec:proposed_approach}

The estimation framework proposed in this paper is an extension of the \ac{BCE} methodology as detailed within~\cite{watson2019enabling}. In this section we have two primary objectives. Firstly, to expound the differentiating factors between the two implementations. Secondly, to provide an overview of the proposed framework entitled \ac{BCE-AD}.

\subsection{The Data Model} \label{subsec:data_model}

\begin{figure}[!htb]
    \begin{subfigure}{\linewidth}
        \centering
        \includegraphics[width=0.85\linewidth]{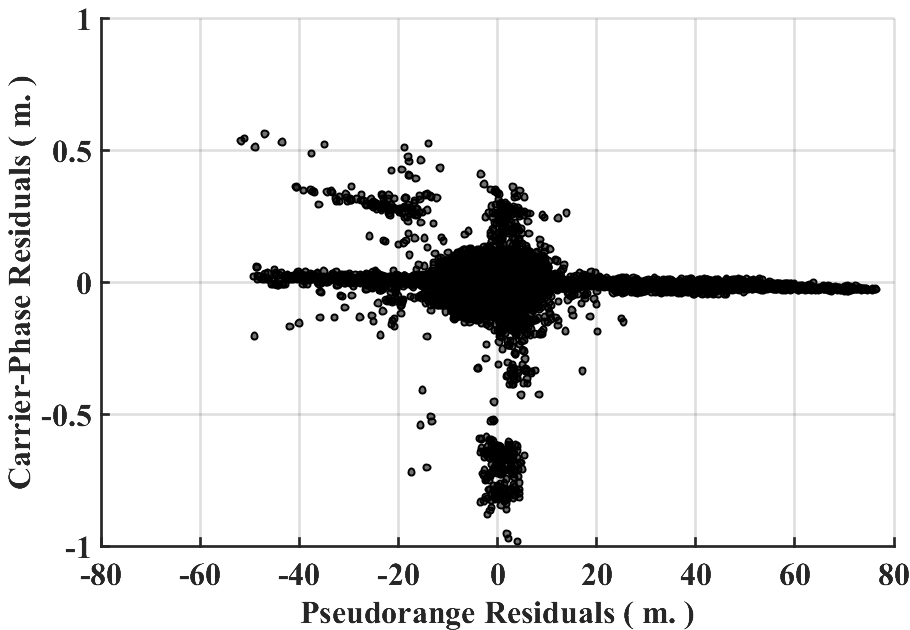} 
         \caption{Measurement residual data domain (i.e., pseudorange and carrier-phase residuals) extracted after the initial iteration of $l^2$ optimization.}
        \label{fig:bce_residuals}
    \end{subfigure}
    \begin{subfigure}{\linewidth}
        \centering
        \includegraphics[width=0.85\linewidth]{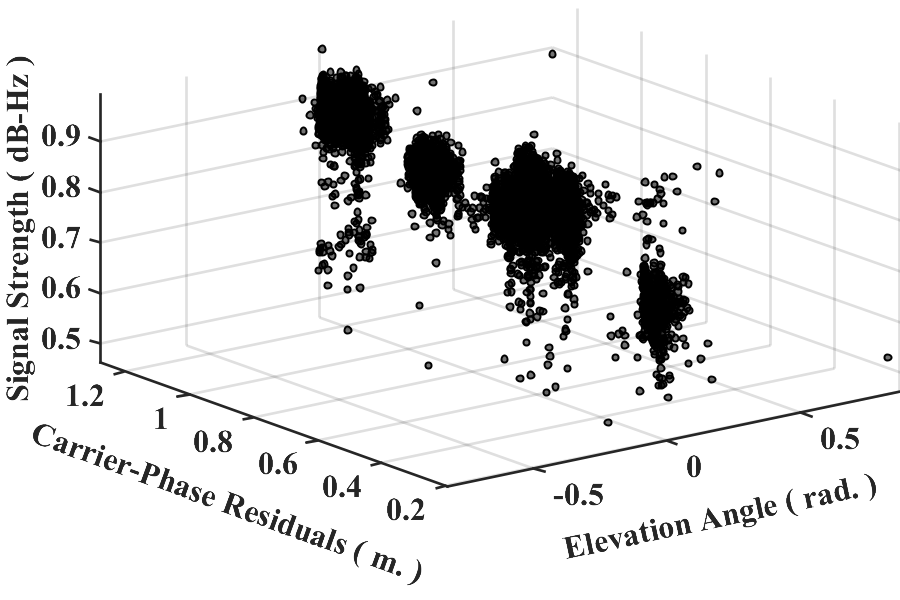}
        \caption{Augmented data domain (i.e., carrier-phase residuals, elevation angle, and signal strength) extracted after the initial iteration of optimization.}
\label{fig:augmented_data}
\end{subfigure}
\caption{Data separability comparison in two domains, where, Fig. \ref{fig:bce_residuals} depicts the measurement residual domain, and Fig. \ref{fig:augmented_data} depicts the augmented data domain.}
    \label{fig:1}
\end{figure}

To initiate a discussion of the proposed estimation framework, we will start with the assumed data model. In our previous work~\cite{watsonbatch,watson2019enabling}, it is assumed that a given set of residuals $R = \{r_1, r_2, \hdots, r_n\} $ with $r_n  = y_n - h_n(X) \in \mathbb{R}^d$ can be accurately partitioned into groupings of similar instances. To depict this visually, Fig. \ref{fig:bce_residuals} provides the \ac{GNSS} measurement residuals (i.e., the pseudorange and carrier-phase residuals) for a typical localization application. As illustrated, there is no obvious partitioning of the set. 

To increase the likelihood of a unique partitioning, we can project the residual dataset into a higher-dimensional space~\cite{cover1965geometrical}. The specific projection utilized within this study is the augmentation of the original dataset with a set of metadata (i.e., for each calculated residual, there is an additional set of features $F = \{f_1, f_2, \hdots, f_n\} $ with $f_n \in \mathbb{R}^f$). Where a feature considered for inclusion in the metadata-set is an observed quantity that is known to correlate to the quality of the collected sensor observation\footnote{For GNSS applications, signal strength and elevation angle of a \ac{GNSS} observation are commonly utilized. Other sensor have similarly useful features (e.g., the mean illumination of an image for computer vision applications)\\}. To depict the benefit granted by the incorporation of metadata, Fig. \ref{fig:augmented_data} provides the carrier-phase residuals augmented by two additional features (i.e., elevation angle and signal strength of the collected observation). From this figure, it can be seen that the inclusion of metadata provides a more obvious partitioning.

Given this set of augmented data $D =\{ d_1, d_2, \hdots, d_n\} $ with $d_n = \{r_n, f_n\} \in \mathbb{R}^{d+f}$. We will assume that the augmented set can be partitioned into similar groupings (i.e., $\bigcup_{m=1}^{M} C_{m} = D$), where each group, $C_m$, can be characterized by a Gaussian distribution (i.e., $C_m \sim \mathcal{N}(\mu_m, \Lambda_m)$). With this assumed model, the augmented dataset is fully characterized as a \ac{GMM}, as depicted in Eq. \ref{eq:res_gmm}

\begin{equation}
 D \sim \sum_{m=1}^M w_m \mathcal{N}(D \ | \ \theta_m) \quad \text{s.t.} \quad \theta_m \triangleq \{ \mu_m, \Lambda_m\},
 \label{eq:res_gmm}
\end{equation}

\noindent where, $m$ is the number of components in the mixture model\footnote{The number of mixture components $M$ in the estimated model is not assumed to be known \apre. Instead, a truncation level $M^{*}$ is set~\cite{kurihara2007accelerated} (i.e., a maximum number of components, $M^{*} \geq M$) and the number of utilized components is autonomously selected based upon the variational free energy of the estimated model~\cite{steinberg2013unsupervised}.\\ }, $w$ is the set of mixture weights with the constraint that $ \sum_m w_m = 1$, and $\theta_m$ is the mixture components sufficient statistics.

\subsection{Variational Clustering} \label{subsec:clustering}

As discussed within~\cite{watson2019enabling}, to fit the a \ac{GMM} to a provided dataset there are two broad classes of algorithms. There are the sampling based frameworks (i.e., Monte Carlo Based approaches~\cite{doucet2005monte}), and the optimization based frameworks (i.e., the variational based approaches~\cite{bishop2006pattern}). Within this study, the variational clustering framework~\cite{bishop2006pattern, steinberg2013unsupervised} is utilized to reduce the computational complexity of the fitting process.

Specifically, the variational clustering approach is utilized to estimate the model parameters that maximize the $\operatorname{log \ marginal \ likelihood}$ of the data, as provided in Eq. \ref{eq:model_fit}

\begin{equation}
 \operatorname{log}\operatorname{p}(D) = \operatorname{log} \int \operatorname{p}(D, \theta, Z) d\mathbf{Z} d\boldsymbol{\theta},
 \label{eq:model_fit}
\end{equation}

\noindent where $Z = \{z_1, z_2, \ldots z_m \}$ with $z_m \in \mathbb{R}^M$ is a set of assignment variables (i.e., Z provides the explicit assignment of each data instance to a component within the \ac{GMM}). To enable the tractable computation of the \ac{GMM}, the mean-field assumption \cite{blei2017variational} (i.e., $\operatorname{p}(D,\theta, Z) \approx \operatorname{q}(\theta)\operatorname{q}(Z)$), is utilized to construct a lower-bound on the true $\operatorname{log \ marginal \ likelihood}$.

With the mean-field assumption in place, the \ac{GMM} parameters can be estimated with the variational framework in a iterative fashion, where the sequence of iteration is presented in Eqs. \ref{eq:vbe_max} and \ref{eq:vbe_exp}. The assignment parameters $Z$ are updated by optimizing Eq. \ref{eq:vbe_max}, where $C_z$ is the normalizing constant to $q(z)$. Then model parameters $\theta$ are updated by holding the assignment parameters fixed and optimizing Eq. \ref{eq:vbe_exp}, where $C_\theta$ is the normalizing constant for $q(\theta)$. This iterative process is continued until the Kullback-Leibler divergence between the approximating and true distributions is minimized.

\begin{equation}
 \operatorname{q}(Z)_{t+1} = C_z \int \operatorname{q}(\theta)_t \operatorname{log} p(D,Z \ | \ \theta) d\theta
 \label{eq:vbe_max}
\end{equation}

\begin{equation}
 \operatorname{q}(\theta)_{t+1} = C_\theta \operatorname{p}(\theta) \int \operatorname{q}(Z)_{t+1} \operatorname{log} \operatorname{p}(D,Z \ | \ \theta) dZ
 \label{eq:vbe_exp}
\end{equation}

\subsection{Feature Selection} \label{subsec:feature_selection}

The inclusion of additional features in the assumed data model have the benefit of increasing the likelihood of a unique partitioning in the measurement domain. However, clustering over an augmented dataset (i.e., a dataset with increased dimensionality) could have the detrimental side effect of increased computation complexity. Thus, to remain computational tractable, the set of augmenting data must be intelligently selected.

The enabling frameworks for intelligent feature selection can be classified as either offline, or online. The offline approach can either be implemented algorithmically of through the utilization of an area expert (i.e., someone with extensive experience with the utilized sensing modality). A primary drawback of the offline feature selection implementation is a static feature space model (i.e., the utilized features are provided \apre and remain fixed over the duration of the estimation process).

Due to the undesirable properties of an offline feature selection approach, an online framework is utilized within this study. Specifically the framework developed within~\cite{cai2010unsupervised} is utilized to autonomously select the most relevant features between each iteration of optimization. In brief, the utilized feature selection algorithm can be described through three primarily steps: 1) construct a \ac{NN} graph from the provided augmented dataset, 2) conduct eigenvalue decomposition on the \ac{NN} graph to measure the importance of each feature for partitioning the dataset, 3) perform \ac{LARS}~\cite{hesterberg2008least} on the calculated eigenvectors, to find the most important features. For a more thorough discussion on the feature selection algorithm, the reader is referred to~\cite{cai2010unsupervised}. 

\subsection{Algorithm Overview}

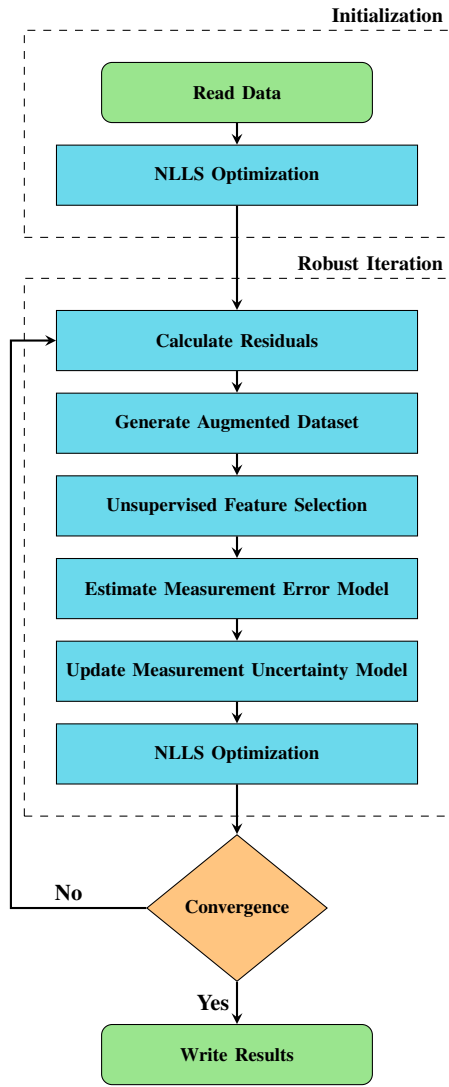
\begin{figure}
 \centering
\begin{tikzpicture}[ node distance=1.1cm]
  \node (start) [startstop, style={scale=0.8}] {Read Data};
  \node (pro1) [process, below of=start, style={scale=0.8}] {NLLS Optimization};
  \node (pro2) [process, below of=pro1, yshift=-1.1cm, style={scale=0.8}] {Calculate Residuals};
  \node (pro2_1) [process, below of=pro2, style={scale=0.8}] {Generate Augmented Dataset};
  \node (pro2_2) [process, below of=pro2_1, style={scale=0.8}] {Unsupervised Feature Selection};
  \node (pro3) [process, below of=pro2_2, style={scale=0.8}] {Estimate Measurement Error Model};
  \node (pro4) [process, below of=pro3, style={scale=0.8}] {Update Measurement Uncertainty Model};
  \node (pro5) [process, below of=pro4, style={scale=0.8}] {NLLS Optimization};
  \node (dec1) [decision, below of=pro5, yshift=-0.95cm, style={scale=0.8}] {Convergence};
  \node (stop) [startstop, below of=dec1, yshift=-0.85cm, style={scale=0.8}] {Write Results};
  \draw [arrow] (start) -- (pro1);
  \draw [arrow] (pro1) -- (pro2);
  \draw [arrow] (pro2) -- (pro2_1);
  \draw [arrow] (pro2_1) -- (pro2_2);
  \draw [arrow] (pro2_2) -- (pro3);
  \draw [arrow] (pro3) -- (pro4);
  \draw [arrow] (pro4) -- (pro5);
  \draw [arrow] (pro5) -- (dec1);
  \draw [arrow] (dec1.west) node[above left, xshift=-0.75cm, style={scale=0.85}]   {\large{\textbf{No}}} -- + (-18mm,0) |- (pro2.west);
  \draw [arrow] (dec1) -- node[anchor=east, style={scale=0.85}] {\large{\textbf{Yes}}} (stop);

  \node (db1) [draw,inner sep=12pt, dashed, fit=(start) (pro1)] {};
  \node[above left, style={scale=0.85}] at (db1.north east) {\textbf{Initialization}};

  \node (db2) [draw,inner sep=12pt, dashed,fit=(pro2) (pro2_1) (pro2_2) (pro3) (pro4) (pro5)] {};
  \node[above left, style={scale=0.85}] at (db2.north east) { {\textbf{Robust Iteration} }};

 \end{tikzpicture}
 \caption{Overview of the proposed robust optimized algorithm. The proposed approach enables robust state estimation through the iterative estimation of the -- possible multimodal -- measurement error covariance model, where the measurement error covariance model is estimated by clustering over an augmented data space constructed from the previous iteration of optimization.}
 \label{fig:algo}
\end{figure}

With the previously detailed topics, the discussion can now proceed to an overview of the proposed estimation framework. From Fig. \ref{fig:algo}, it is shown that the proposed algorithm is comprised of two primary segments. The first component is the initialization of the estimator. This process begins by constructing the factor graph representation of the \ac{NLLS} optimization problem -- as thoroughly detailed within~\cite{dellaert2017factor} -- from the {\textit{a priori}} state and uncertainty information and the provided observations. Then, an initial iteration of optimization is conducted with a \ac{NLLS} estimation algorithm (e.g., for this study, the Levenberg-Marquardt~\cite{more1978levenberg} implementation was utilized) to update the \apre state estimate.

The second component of the proposed framework commences with the calculation of the measurement residuals given the previously estimated set of states. The set of calculated residuals are augmented with the provided set of metadata, as discussed in section \ref{subsec:data_model}. Using the augmented dataset for the current iteration of optimization, the most relevant features are selected using the framework discussed in section \ref{subsec:feature_selection}. Using the relevant set of features, the variational cluster framework -- as disused in section \ref{subsec:clustering} -- is utilized to assign each instance in the augmented dataset to a component within an estimated \ac{GMM}. 

Utilizing the partitioning of the augmented data space, the measurement error uncertainty model can be calculated. This is achieved by first partitioning the measurement residuals with the assignment vector $Z$ estimated in the augmented data space. Then, the sufficient statistics for each grouping in the measurement residual domain are calculated. With this $m\text{-dimensional}$ \ac{GMM} in the measurement residual domain, the measurement uncertainty model of the factor graph is updated (i.e., each measurement's uncertainty model is updated to the sufficient statistics of the assigned \ac{GMM} component). With the updated measurement uncertainty model, a new iteration of optimization is conducted. This process is iterated until a measure of convergence (e.g., the error decrease between consecutive iterations is less than a user defined threshold) -- or a limit on the number of iterations -- has been reached.

\section{Experimental Validation} \label{sec:experimental_validation}

\begin{figure*}
\centering
    \begin{subfigure}{0.32\linewidth}
\includegraphics[width=\linewidth]{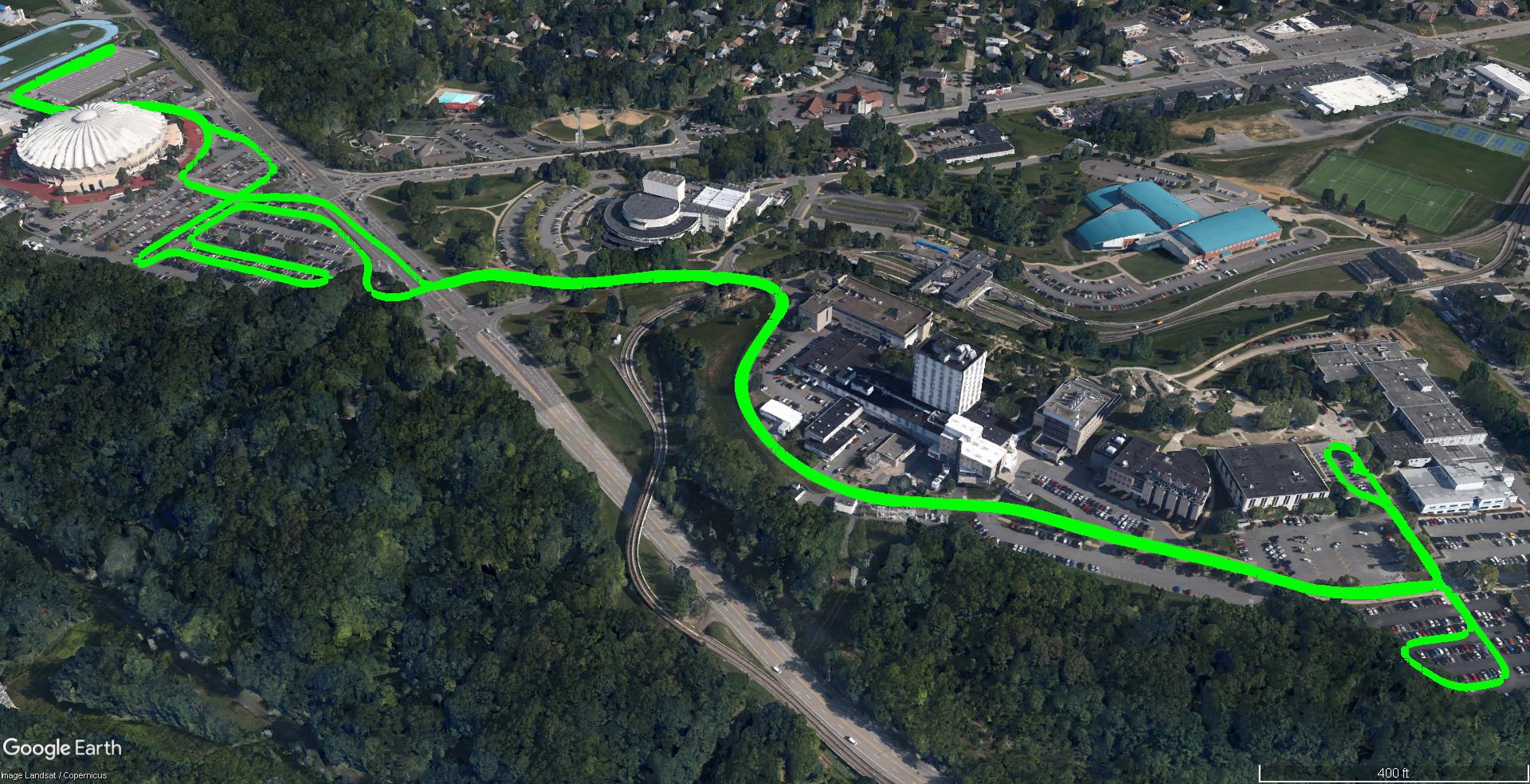} 
    \caption{Ground trace for data collect 1.}
\label{fig:gt_1}
    \end{subfigure} \hfill
    \begin{subfigure}{0.32\linewidth}
\includegraphics[width=\linewidth]{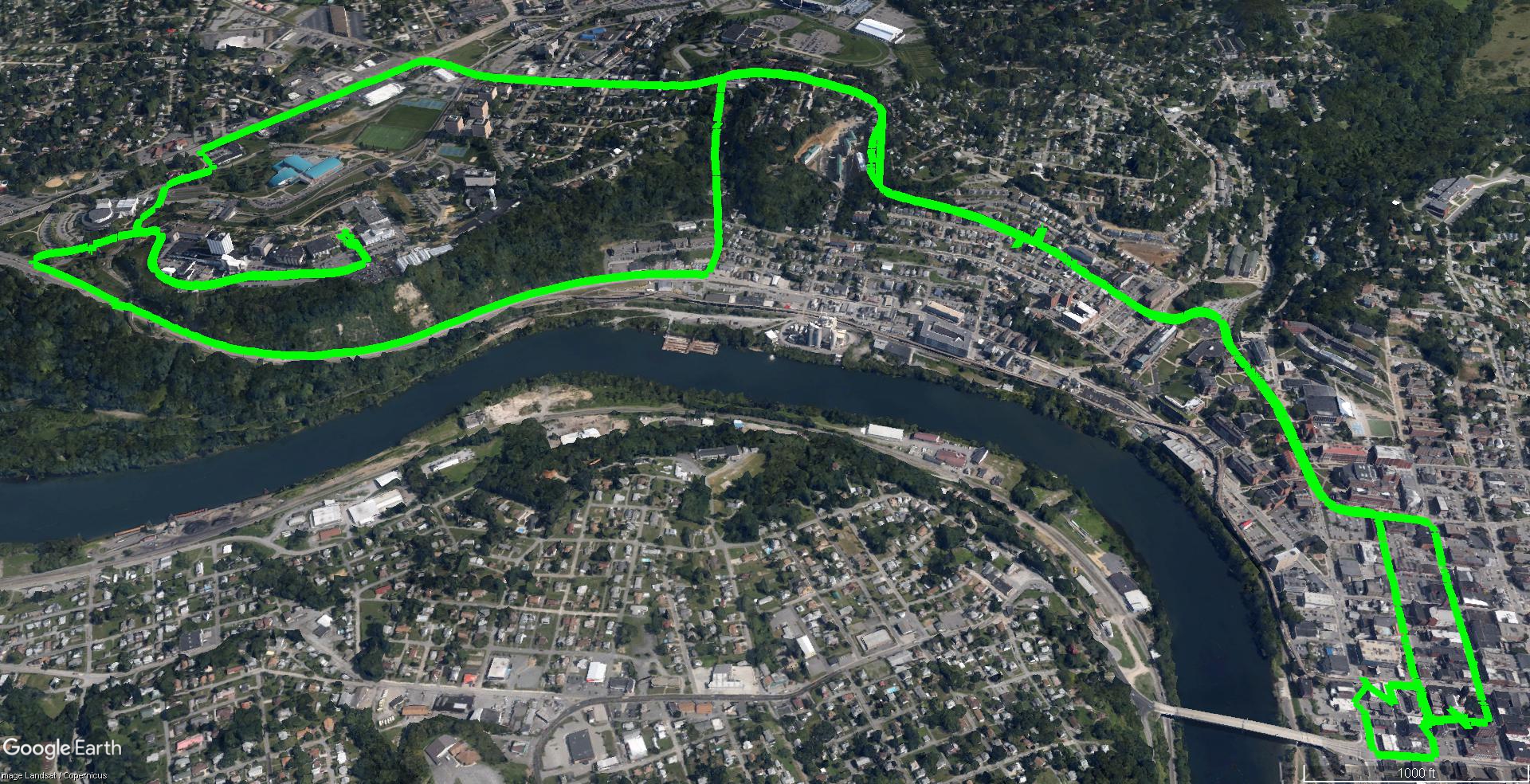}
    \caption{Ground trace for data collect 2.}
\label{fig:gt_2}
    \end{subfigure} \hfill
    \begin{subfigure}{0.32\linewidth}
\includegraphics[width=\linewidth]{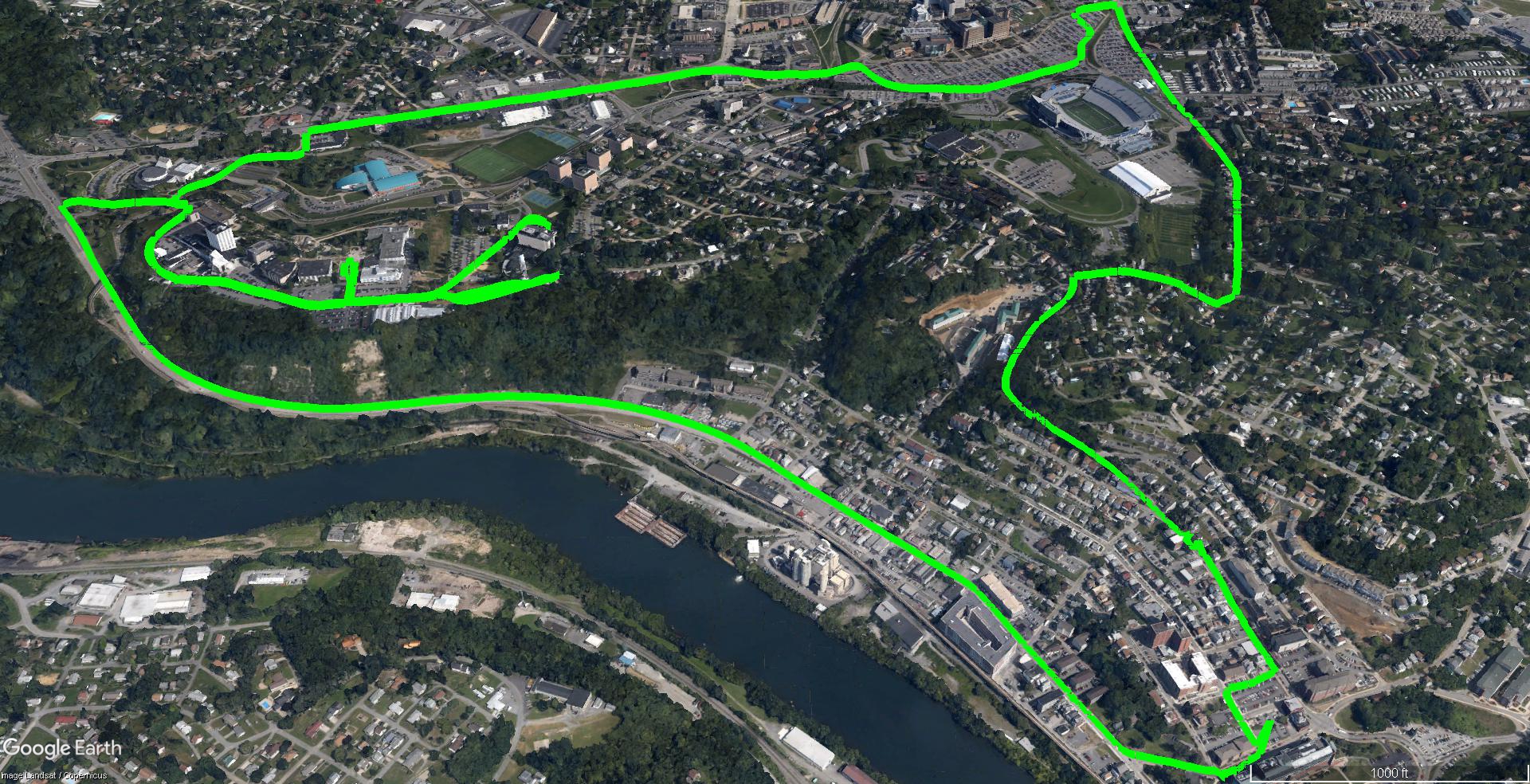}
    \caption{Ground trace for data collect 3.}
    \label{fig:gt_3}
    \end{subfigure}
    \caption{Ground trace for the three kinematic GNSS data collects, which were made publicly available within~\cite{watson2019enabling}. All three data collected were recorded in Morgantown, WV.}
    \label{fig:ground_traces}
\end{figure*}

\subsection{Data Collection}

To enable the validation of the proposed estimation framework, three kinematic \ac{GNSS} data collects\footnote{These three \ac{GNSS} data collects were made publicly available within~\cite{watson2019enabling}. For a detailed description of the data collection procedure, the reader is referred to Section IV of~\cite{watson2019enabling}} are utilized. The ground-trace for the three driving data collects is provided in Fig. \ref{fig:ground_traces}. For each of the depicted data collects, the \ac{GNSS} binary \ac{IQ} observations were recorded with a LabSat-3 GPS record and playback device \cite{labSat3}. The collected \ac{IQ} data were then played back into two \ac{GNSS} receivers: a geodetic-grade (Novatel OEM-638) \ac{GNSS} receiver, and an open-source \ac{GPS} \ac{SDR}.

The observations generated by the geodetic-grade receiver, in conjunction with the observations generated by an additional static \ac{GNSS} receiver, were utilized to calculate a reference positioning solution for each data collect. The reference solution was calculated through a differential (i.e., \ac{RTK}) filter-smoother framework implemented within the open-source package RTKLIB \cite{takasu2011rtklib}. 

The observations generated by the \ac{GPS} \ac{SDR} are utilized by the estimation framework to validate the proposed methodology. These observations are utilized for validation because they were intentionally degraded by altering the \ac{GNSS} receiver's tracking parameters. For a more detailed discussion on the specific \ac{GNSS} tracking parameter settings, the reader is referred to Section IV of~\cite{watson2019enabling}. 

With these three datasets, the proposed approach (i.e., \ac{BCE-AD}) is validated against four additional estimators. The first comparison algorithm is a traditional (i.e., non-robust) estimation framework that utilizes a $l^2\text{-norm}$ cost function. Additionally, the \ac{DCS} robust estimation framework~\cite{DCS} is utilized. The third comparison algorithm is the \ac{MM} estimation framework~\cite{maxmix} with a static measurement uncertainty model. The final comparison algorithm is the \ac{BCE} approach~\cite{watson2019enabling}.  

To enable the utilization of the \ac{BCE-AD} approach, the set of features $F$ need to be specified. For \ac{GNSS} applications, several useful metadata can be calculated for each observation. For this evaluation, it was elected to use three such features: the \ac{SS}, the \ac{EL}, and the \ac{AZ}. These features were selected for two primary reasons: 1) these features can be easily extracted for each \ac{GNSS} observation, 2) these features are known to correlate to the quality of the recorded \ac{GNSS} signal.

\subsection{Results}

\begin{figure}[H]
    \centering
    \includegraphics[width=0.85\columnwidth]{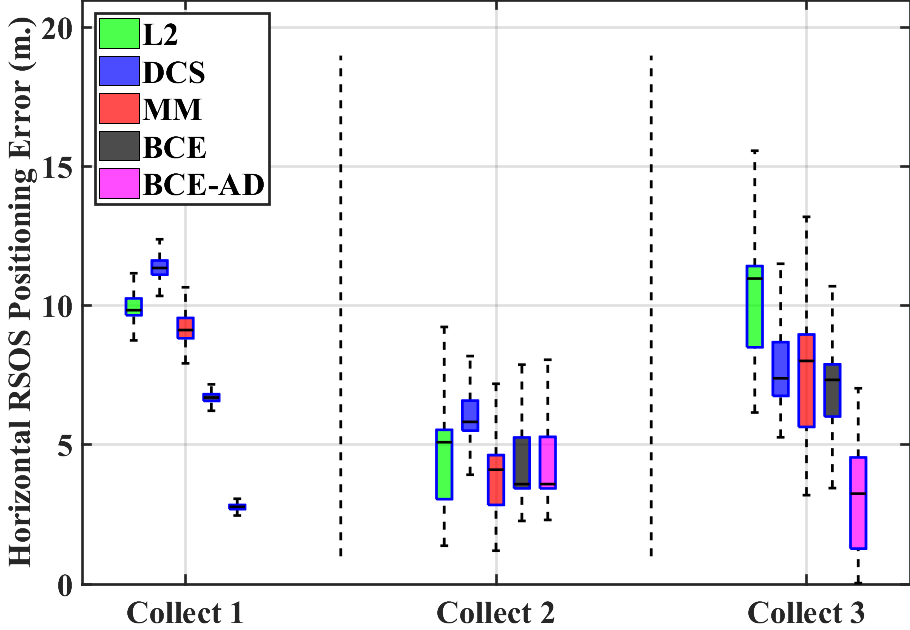}
    \caption{Horizontal RSOS positioning error for the collected GNSS data sets. Within this figure, $L_2$, is a batch estimator with $l^2$ cost function, DCS is the dynamic covariance scaling robust estimator, MM is the max-mixtures approach with a static measurement covariance model, BCE is the batch covariance estimation technique, and BCE-AD is the proposed batch covariance estimation over an augmented data-space. }
    \label{fig:lq_boxplot}
\end{figure}

To begin the evaluation of the proposed methodology, the horizontal \ac{RSOS} positioning error for the three data collects will be examined. This evaluation is presented visually in Fig. \ref{fig:lq_boxplot}, in the form of a box-plot. From Fig. \ref{fig:lq_boxplot} it is shown that the proposed \ac{BCE-AD} significantly reduces the median horizontal \ac{RSOS} positioning error for data collects 1 and 3, and performs comparably well to the other robust estimators on data collect 2. The specific statistics representing the analysis presented in Fig. \ref{fig:lq_boxplot} are provided in Table \ref{table:lq_stats}. 

\begin{table}[H]
 \caption{Horizontal RSOS error statistics with green and red entries corresponding to the minimum or maximum statistic, respectively.}
 \begin{subtable}{1.0\linewidth}
  \centering
  \caption{Horizontal RSOS positioning error results for data collect 1.}
  \resizebox{0.95\columnwidth}{!}{%
  \begin{tabular}{||l|c|c|c|c|c||}
  \hline
  (m.)     & $L_2$ & DCS                       & MM                       & BCE    & BCE-AD                     \\
  \hline\hline

  median   & 9.84  & \cellcolor{red!60}{10.82} & 9.13                     & 6.70  & \cellcolor{green!60}{2.14} \\
  \hline
  variance & 0.33  & 0.21                      & \cellcolor{red!60}{0.37} & \cellcolor{green!60}{0.09}  & 0.16 \\
  \hline
  max      & 14.84 & \cellcolor{red!60}{16.10} & 14.11                    & 11.75 & \cellcolor{green!60}{7.79} \\
  \hline
  \end{tabular}%
  }
  \label{table:drive_1_lq}
 \end{subtable}

 \begin{subtable}{1.0\linewidth}
  \centering

  \vspace{1em}
  \caption{Horizontal RSOS positioning error results for data collect 2.}
  \resizebox{0.95\columnwidth}{!}{%
  \begin{tabular}{||l|c|c|c|c|c||}
  \hline
  (m.)     & $L_2$                    & DCS                          & MM                         & BCE     & BCE-AD                    \\ 
  \hline\hline

  median   & \cellcolor{red!60}{5.09} & 5.02                         & 4.11                       & \cellcolor{green!60}{3.58} & \cellcolor{green!60}{3.58} \\
  \hline
  variance & 613.13                   & 342.92 & \cellcolor{red!60}{673.50} & \cellcolor{green!60}{393.32} & \cellcolor{green!60}{393.32}                \\
  \hline
  max      & 127.29                   & 98.05  & \cellcolor{red!60}{132.49} & \cellcolor{green!60}{103.64} & \cellcolor{green!60}{103.64}                     \\
  \hline
  \end{tabular}%
  }
  \label{table:drive_2_lq}
 \end{subtable}
 \begin{subtable}{1.0\linewidth}
  \centering
  \vspace{1em}
  \caption{Horizontal RSOS positioning error results for data collect 3.}
  \resizebox{0.95\columnwidth}{!}{%
  \begin{tabular}{||l|c|c|c|c|c||}
  \hline
  (m.)     & $L_2$                     & DCS                        & MM                          & BCE   & BCE-AD                     \\
  \hline\hline

  median   & \cellcolor{red!60}{10.98} & 9.88                       & 8.02                        & 7.31  & \cellcolor{green!60}{3.25}\\
  \hline
  variance & 3.06                      & \cellcolor{green!60}{1.49} & \cellcolor{red!60}{3.36}    & 2.55  & 3.25                     \\
  \hline
  max      & 17.26                     & \cellcolor{red!60}{17.30}  & 15.06 & 15.35 & \cellcolor{green!60}{12.64}                      \\
  \hline
  \end{tabular}%
  }
  \label{table:drive_3_lq}
 \end{subtable}
 \label{table:lq_stats}
\end{table}

\begin{figure}
    \centering
    \begin{subfigure}{0.45\linewidth}
        \includegraphics[width=\linewidth]{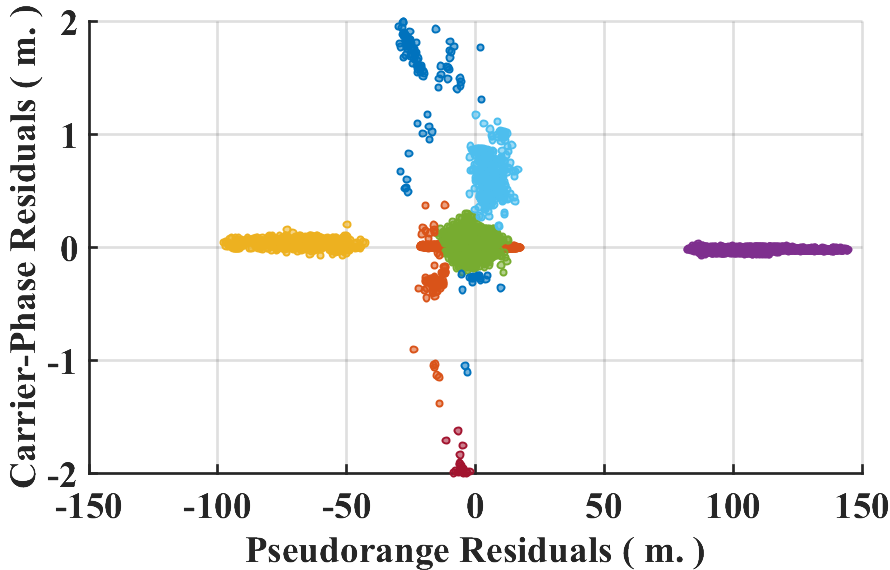} 
        \caption{\ac{BCE} based measurement residual domain partitioning for data collect 1.}
        \label{fig:bce_assign_1}
    \end{subfigure}\hfill
    \begin{subfigure}{0.45\linewidth}
        \includegraphics[width=\linewidth]{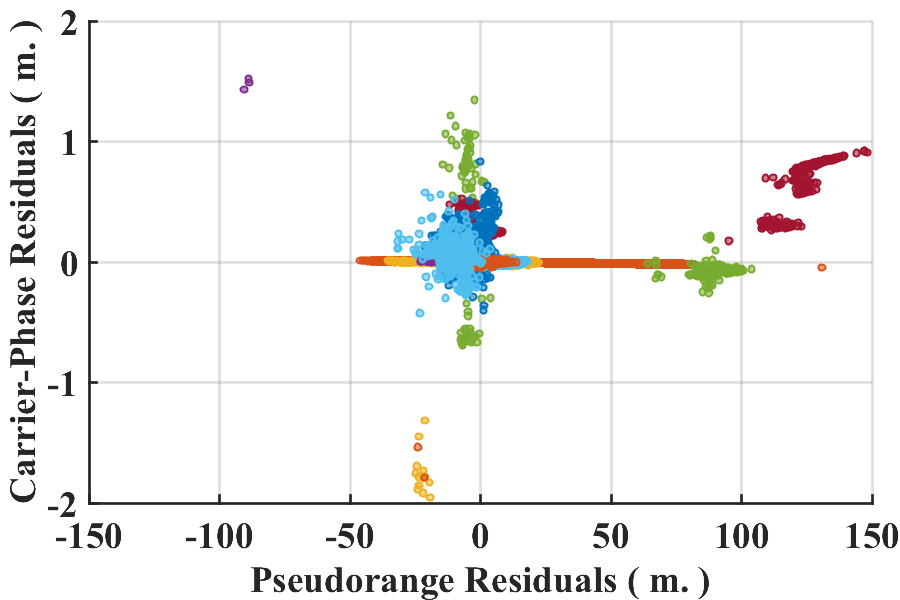}
        \caption{\ac{BCE-AD} based measurement residual domain partitioning for data collect 1.}
        \label{fig:bce_ad_assign_1}
    \end{subfigure}
    
    \begin{subfigure}{0.45\linewidth}
        \includegraphics[width=\linewidth]{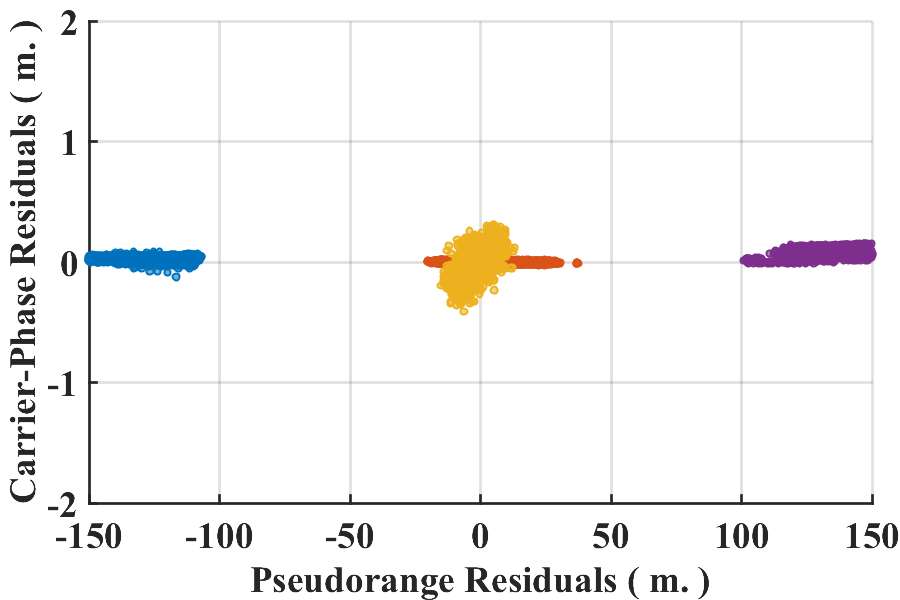}
        \caption{\ac{BCE} based measurement residual domain partitioning for data collect 2.}
        \label{fig:bce_assign_2}
    \end{subfigure}\hfill
    \begin{subfigure}{0.45\linewidth}
        \includegraphics[width=\linewidth]{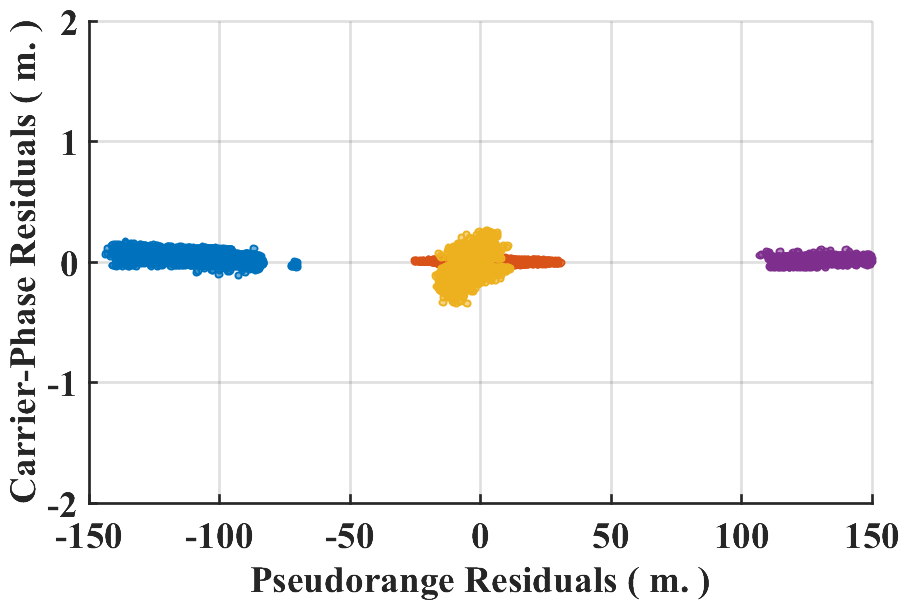}
        \caption{\ac{BCE-AD} based measurement residual domain partitioning for data collect 2.}
        \label{fig:bce_ad_assign_2}
    \end{subfigure}
    
    \begin{subfigure}{0.45\linewidth}
        \includegraphics[width=\linewidth]{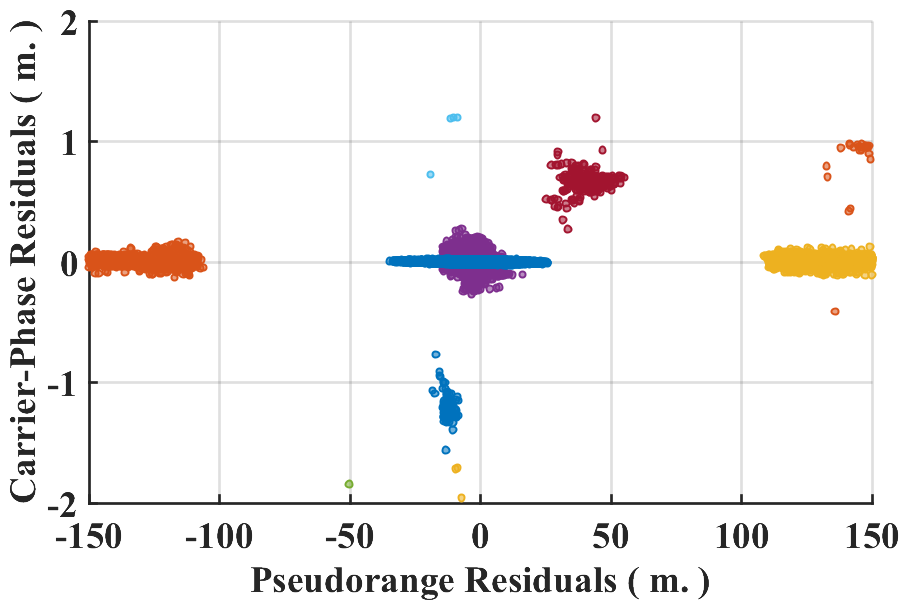}
        \caption{\ac{BCE} based measurement residual domain partitioning for data collect 3.}
        \label{fig:bce_assign_3}
    \end{subfigure}\hfill
    \begin{subfigure}{0.45\linewidth}
        \includegraphics[width=\linewidth]{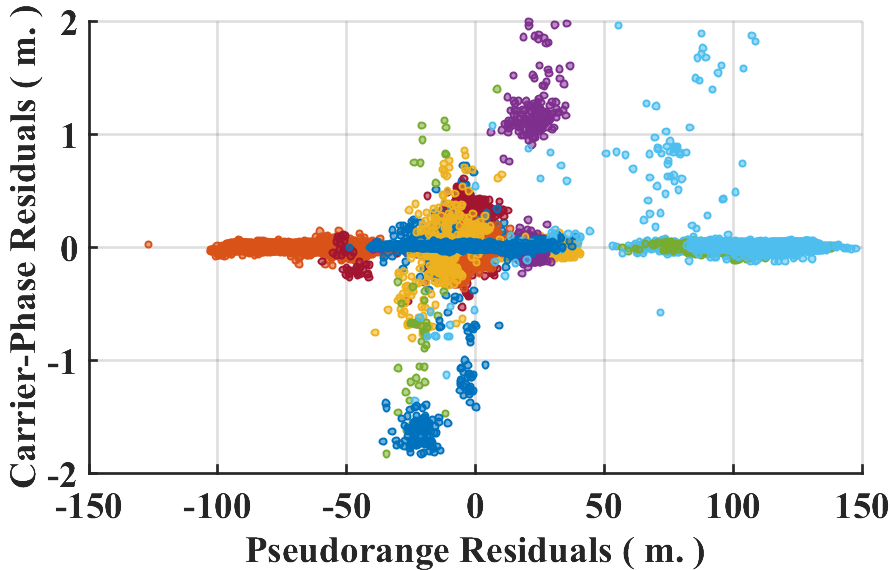}
        \caption{\ac{BCE-AD} based measurement residual domain partitioning for data collect 3.}
        \label{fig:bce_ad_assign_3}
    \end{subfigure}
    \caption{Partitioning of the measurement residual domain with the \ac{BCE} approach (see Figs. \ref{fig:bce_assign_1}, \ref{fig:bce_assign_2}, and \ref{fig:bce_assign_3}), and the \ac{BCE-AD} approach (see Figs. \ref{fig:bce_ad_assign_1}, \ref{fig:bce_ad_assign_2}, and \ref{fig:bce_ad_assign_3}), for the three kinematic GNSS data collects at the final iteration of optimization.}
    \label{fig:assignment}
\end{figure}

To examine the reason why the \ac{BCE-AD} approach provided no additional positioning performance benefit when compared to the \ac{BCE} approach for data collect 2, the measurement domain partitioning for the two approaches can be evaluated. This evaluation is provided in Fig. \ref{fig:assignment} for all three data collects. From the provided figure, it can be hypothesized that no additional positioning accuracy is granted when the measurement residuals can be easily partitioned utilizing exclusively the information in the measurement residual domain. This is depicted visually by evaluating Fig. \ref{fig:bce_assign_2}, where it is shown that the clustering utilizing only the information in the measurement domain can easily partition the provided measurement residuals. However, when evaluating the measurement residuals for data collects 1 and 3 (i.e., the datasets where the greatest positioning performance benefit is granted by the \ac{BCE-AD} approach), it is apparent that clustering over the two different domains provides significantly different partitionings.

 Finally, the utilized features with the \ac{BCE-AD} approach for each dataset are evaluated. This evaluation is depicted visually in Table \ref{table:utilized_features}. From this visual, it can be noted that the most relevant features not only vary from dataset to dataset, but also from one optimization iteration to the next within a given dataset. Additionally, the hypothesis presented in the previous paragraph (i.e., that the \ac{BCE-AD} only grants increased positioning performance when sufficient information is not present in the measurement residual domain to partition the residuals) is further verified in Table \ref{table:features_ds_2} where it is shown that only features utilized for data collect 2 are the measurement residuals.

\begin{table}[htbp]
 \caption{Utilized features for each iteration of optimization with the \ac{BCE-AD} approach. The green and red cell entries correspond to the utilized and non-utilized features, respectively. Within the table, $\rho$ is the pseudorange residual, $\Phi$ is the carrier-phase residual, El. is the elevation angle, Az. is the azimuth angle, and SS. is the signal strength.}
 \begin{subtable}{1.0\linewidth}
  \centering
  \caption{Utilized features per iteration of optimization for data collect 1.}
  \resizebox{0.95\columnwidth}{!}{%
  \begin{tabular}{||l|c|c|c|c|c|c|c|c|c||}
   \hline
      & Iter 0. & Iter 1. & Iter 2.   & Iter 3. & Iter 4.  & Iter 5.   & Iter 6.  & Iter 7. & Iter 8. \\
   \hline\hline

   $\rho$  & \cellcolor{green!60}  & \cellcolor{red!60}  & \cellcolor{green!60} & \cellcolor{green!60} & \cellcolor{green!60} & \cellcolor{green!60} & \cellcolor{green!60} & \cellcolor{green!60} &  \cellcolor{green!60}  \\ \hline
   $\Phi$ & \cellcolor{green!60} & \cellcolor{green!60} & \cellcolor{green!60} & \cellcolor{green!60} & \cellcolor{green!60} & \cellcolor{green!60} & \cellcolor{green!60}  & \cellcolor{green!60} & \cellcolor{green!60}\\ \hline
   El. & \cellcolor{red!60} & \cellcolor{green!60} & \cellcolor{red!60} & \cellcolor{red!60} & \cellcolor{green!60} & \cellcolor{green!60} & \cellcolor{green!60}  & \cellcolor{green!60} & \cellcolor{green!60} \\ \hline
   Az.  & \cellcolor{red!60} & \cellcolor{red!60} & \cellcolor{red!60} & \cellcolor{red!60} & \cellcolor{red!60} & \cellcolor{red!60} & \cellcolor{red!60}  & \cellcolor{red!60} & \cellcolor{red!60} \\ \hline
   SS. & \cellcolor{red!60} & \cellcolor{green!60} & \cellcolor{red!60} & \cellcolor{red!60} & \cellcolor{green!60} & \cellcolor{green!60} & \cellcolor{green!60}  & \cellcolor{green!60} & \cellcolor{green!60} \\ \hline
  \end{tabular}%
  }
  \label{table:features_ds_1}
 \end{subtable}

 \begin{subtable}{1.0\linewidth}
  \centering

  \vspace{1em}
  \caption{Utilized features per iteration of optimization for data collect 2.}
  \resizebox{0.8\columnwidth}{!}{%
  \begin{tabular}{||l|c|c|c|c|c|c|c||}
   \hline
      & Iter 0. & Iter 1. & Iter 2.   & Iter 3. & Iter 4.  & Iter 5.   & Iter 6. \\
   \hline\hline

   $\rho$  & \cellcolor{green!60}  & \cellcolor{green!60}  & \cellcolor{green!60} & \cellcolor{green!60} & \cellcolor{green!60} & \cellcolor{green!60} & \cellcolor{green!60} \\ \hline
   $\Phi$ & \cellcolor{green!60} & \cellcolor{green!60} & \cellcolor{green!60} & \cellcolor{green!60} & \cellcolor{green!60} & \cellcolor{green!60} & \cellcolor{green!60} \\ \hline
   El. & \cellcolor{red!60} & \cellcolor{red!60} & \cellcolor{red!60} & \cellcolor{red!60} & \cellcolor{red!60} & \cellcolor{red!60} & \cellcolor{red!60} \\ \hline
   Az.  & \cellcolor{red!60} & \cellcolor{red!60} & \cellcolor{red!60} & \cellcolor{red!60} & \cellcolor{red!60} & \cellcolor{red!60} & \cellcolor{red!60}  \\ \hline
   SS. & \cellcolor{red!60} & \cellcolor{red!60} & \cellcolor{red!60} & \cellcolor{red!60} & \cellcolor{red!60} & \cellcolor{red!60} & \cellcolor{red!60} \\ \hline
  \end{tabular}%
  }
  \label{table:features_ds_2}
 \end{subtable}
 \begin{subtable}{1.0\linewidth}
  \centering
  \vspace{1em}
  \caption{Utilized features per iteration of optimization for data collect 3.}
  \resizebox{0.65\columnwidth}{!}{%
  \begin{tabular}{||l|c|c|c|c|c||}
   \hline
      & Iter 0. & Iter 1. & Iter 2.   & Iter 3. & Iter 4. \\
   \hline\hline

   $\rho$  & \cellcolor{green!60}  & \cellcolor{green!60}  & \cellcolor{green!60} & \cellcolor{green!60} & \cellcolor{green!60}  \\ \hline
   $\Phi$ & \cellcolor{green!60} & \cellcolor{green!60} & \cellcolor{green!60} & \cellcolor{green!60} & \cellcolor{green!60} \\ \hline
   El. & \cellcolor{red!60} & \cellcolor{red!60} & \cellcolor{green!60} & \cellcolor{red!60} & \cellcolor{green!60}\\ \hline
   Az.  & \cellcolor{red!60} & \cellcolor{red!60} & \cellcolor{red!60} & \cellcolor{red!60} & \cellcolor{red!60}  \\ \hline
   SS. & \cellcolor{red!60} & \cellcolor{green!60} & \cellcolor{green!60} & \cellcolor{green!60} & \cellcolor{green!60} \\ \hline
  \end{tabular}%
  }
  \label{table:featues_ds_3} 
  \end{subtable}
 \label{table:utilized_features}
\end{table}
 
\pagebreak
\section{Conclusion} \label{sec:conclusion}

This paper presents an extension of the previously proposed batch covariance estimation (BCE) technique to enable robust state estimation. The BCE approach enables robust state estimation through the iterative estimation of a measurement error uncertainty model based upon the previous iterations measurement residuals. Where, the estimated measurement error uncertainty model is characterized by a Gaussian mixture model (GMM) which is fit to the measurement residuals through variational clustering. After fitting the GMM to the current optimization iterations residuals, the uncertainty model of each observation is updated to the sufficient statistics of the assigned cluster within the GMM.

The approach proposed within this work extends the BCE approach on one front. Specifically, it removes the assumption that the measurement error uncertainty model can be accurately characterized utilizing information exclusively contained within the measurement residual domain. Instead, this paper argues that the uncertainty estimation process should be augmented to include additional metadata. The modification of the uncertainty estimation process to an augmented data space increases the likelihood of a unique partitioning in the measurement residual domain and thus provides the ability to more accurately characterize the measurement uncertainty model.

To verify the proposed batch covariance estimation over and augmented data space (BCE-AD) approach, three GNSS data sets were utilized. The utilized data sets provide varying levels of degradation to quantify the robustness of the proposed algorithm against other state-of-the-art robust estimators. Utilizing these data sets, it is shown that the proposed approach provides comparable or improved state estimation accuracy when compared to other robust estimation techniques. In conclusion, it should be noted that, while the approach was validated on \ac{GNSS} observations for this study, the proposed approach is generically extensible to other estimation domains. 
\section*{Acknowledgment}

The authors would like to thank MacAuly-Brown Inc. for funding the work presented within this article.

\bibliography{main}

\begin{thebibliography}{10}
\providecommand{\url}[1]{#1}
\csname url@samestyle\endcsname
\providecommand{\newblock}{\relax}
\providecommand{\bibinfo}[2]{#2}
\providecommand{\BIBentrySTDinterwordspacing}{\spaceskip=0pt\relax}
\providecommand{\BIBentryALTinterwordstretchfactor}{4}
\providecommand{\BIBentryALTinterwordspacing}{\spaceskip=\fontdimen2\font plus
\BIBentryALTinterwordstretchfactor\fontdimen3\font minus
  \fontdimen4\font\relax}
\providecommand{\BIBforeignlanguage}[2]{{%
\expandafter\ifx\csname l@#1\endcsname\relax
\typeout{** WARNING: IEEEtran.bst: No hyphenation pattern has been}%
\typeout{** loaded for the language `#1'. Using the pattern for}%
\typeout{** the default language instead.}%
\else
\language=\csname l@#1\endcsname
\fi
#2}}
\providecommand{\BIBdecl}{\relax}
\BIBdecl

\bibitem{mallios2016toward}
A.~Mallios, P.~Ridao, D.~Ribas, M.~Carreras, and R.~Camilli, ``Toward
  autonomous exploration in confined underwater environments,'' \emph{Journal
  of Field Robotics}, vol.~33, no.~7, pp. 994--1012, 2016.

\bibitem{barfoot2017state}
T.~D. Barfoot, \emph{State Estimation for Robotics}.\hskip 1em plus 0.5em minus
  0.4em\relax Cambridge University Press, 2017.

\bibitem{simon2006optimal}
D.~Simon, \emph{{Optimal state estimation: Kalman, H infinity, and nonlinear
  approaches}}.\hskip 1em plus 0.5em minus 0.4em\relax John Wiley \& Sons,
  2006.

\bibitem{thrun2005probabilistic}
S.~Thrun, W.~Burgard, and D.~Fox, \emph{Probabilistic robotics}.\hskip 1em plus
  0.5em minus 0.4em\relax MIT press, 2005.

\bibitem{dellaert2017factor}
F.~Dellaert, M.~Kaess \emph{et~al.}, ``Factor graphs for robot perception,''
  \emph{Foundations and Trends{\textregistered} in Robotics}, vol.~6, no. 1-2,
  pp. 1--139, 2017.

\bibitem{hampel1968contribution}
F.~R. Hampel, ``Contribution to the theory of robust estimation,'' \emph{Ph. D.
  Thesis, University of California, Berkeley}, 1968.

\bibitem{huberBook}
P.~J. Huber, \emph{{Robust Statistics}}.\hskip 1em plus 0.5em minus 0.4em\relax
  Wiley New York, 1981.

\bibitem{DCS}
P.~Agarwal, G.~D. Tipaldi, L.~Spinello, C.~Stachniss, and W.~Burgard, ``Robust
  map optimization using dynamic covariance scaling,'' in \emph{2013 IEEE
  International Conference on Robotics and Automation}.\hskip 1em plus 0.5em
  minus 0.4em\relax Citeseer, 2013, pp. 62--69.

\bibitem{maxmix}
E.~Olson and P.~Agarwal, ``Inference on networks of mixtures for robust robot
  mapping,'' \emph{The International Journal of Robotics Research}, vol.~32,
  no.~7, pp. 826--840, 2013.

\bibitem{fischler1981random}
M.~A. Fischler and R.~C. Bolles, ``Random sample consensus: a paradigm for
  model fitting with applications to image analysis and automated
  cartography,'' \emph{Communications of the ACM}, vol.~24, no.~6, pp.
  381--395, 1981.

\bibitem{zhai2018fault}
Y.~Zhai, M.~Joerger, and B.~Pervan, ``Fault exclusion in multi-constellation
  global navigation satellite systems,'' \emph{The Journal of Navigation},
  vol.~71, no.~6, pp. 1281--1298, 2018.

\bibitem{carlone2014selecting}
L.~Carlone, A.~Censi, and F.~Dellaert, ``{Selecting good measurements via $l^1$
  relaxation: A convex approach for robust estimation over graphs},'' in
  \emph{2014 IEEE/RSJ International Conference on Intelligent Robots and
  Systems}.\hskip 1em plus 0.5em minus 0.4em\relax IEEE, 2014, pp. 2667--2674.

\bibitem{watsonbatch}
R.~M. Watson, C.~N. Taylor, R.~C. Leishman, and J.~N. Gross, ``{Batch
  Measurement Error Covariance Estimation for Robust Localization},'' in
  \emph{ION GNSS+ 2018}.\hskip 1em plus 0.5em minus 0.4em\relax the Institute
  of Navigation, 2018, pp. 2429--2439.

\bibitem{watson2019enabling}
R.~M. Watson, J.~N. Gross, C.~N. Taylor, and R.~C. Leishman, ``{Enabling Robust
  State Estimation through Measurement Error Covariance Adaptation},''
  \emph{arXiv preprint arXiv:1906.04055}, 2019.

\bibitem{dellaert2006square}
F.~Dellaert and M.~Kaess, ``{Square Root SAM: Simultaneous localization and
  mapping via square root information smoothing},'' \emph{The International
  Journal of Robotics Research}, vol.~25, no.~12, pp. 1181--1203, 2006.

\bibitem{zhang1997parameter}
Z.~Zhang, ``Parameter estimation techniques: A tutorial with application to
  conic fitting,'' \emph{Image and vision Computing}, vol.~15, no.~1, pp.
  59--76, 1997.

\bibitem{pfeifer2018expectation}
T.~Pfeifer and P.~Protzel, ``Expectation-maximization for adaptive mixture
  models in graph optimization,'' \emph{arXiv preprint arXiv:1811.04748}, 2018.

\bibitem{cover1965geometrical}
T.~M. Cover, ``Geometrical and statistical properties of systems of linear
  inequalities with applications in pattern recognition,'' \emph{IEEE
  transactions on electronic computers}, no.~3, pp. 326--334, 1965.

\bibitem{kurihara2007accelerated}
K.~Kurihara, M.~Welling, and N.~Vlassis, ``Accelerated variational dirichlet
  process mixtures,'' in \emph{Advances in neural information processing
  systems}, 2007, pp. 761--768.

\bibitem{steinberg2013unsupervised}
D.~Steinberg, ``An unsupervised approach to modelling visual data,'' Ph.D.
  dissertation, University of Sydney., 2013.

\bibitem{doucet2005monte}
A.~Doucet and X.~Wang, ``{Monte Carlo methods for signal processing: a review
  in the statistical signal processing context},'' \emph{IEEE Signal Processing
  Magazine}, vol.~22, no.~6, pp. 152--170, 2005.

\bibitem{bishop2006pattern}
C.~M. Bishop, \emph{Pattern recognition and machine learning}.\hskip 1em plus
  0.5em minus 0.4em\relax springer, 2006.

\bibitem{blei2017variational}
D.~M. Blei, A.~Kucukelbir, and J.~D. McAuliffe, ``Variational inference: A
  review for statisticians,'' \emph{Journal of the American Statistical
  Association}, vol. 112, no. 518, pp. 859--877, 2017.

\bibitem{cai2010unsupervised}
D.~Cai, C.~Zhang, and X.~He, ``Unsupervised feature selection for multi-cluster
  data,'' in \emph{Proceedings of the 16th ACM SIGKDD international conference
  on Knowledge discovery and data mining}.\hskip 1em plus 0.5em minus
  0.4em\relax ACM, 2010, pp. 333--342.

\bibitem{hesterberg2008least}
T.~Hesterberg, N.~H. Choi, L.~Meier, C.~Fraley \emph{et~al.}, ``Least angle and
  $l^1$ penalized regression: A review,'' \emph{Statistics Surveys}, vol.~2,
  pp. 61--93, 2008.

\bibitem{more1978levenberg}
J.~J. Mor{\'e}, ``The levenberg-marquardt algorithm: implementation and
  theory,'' in \emph{Numerical analysis}.\hskip 1em plus 0.5em minus
  0.4em\relax Springer, 1978, pp. 105--116.

\bibitem{labSat3}
``{LabSat 3 GPS Simulator},''
  \url{https://www.labsat.co.uk/index.php/en/products/labsat-3}, accessed
  3/1/19.

\bibitem{takasu2011rtklib}
T.~Takasu, ``{RTKLIB: An open source program package for GNSS positioning},''
  2011.

\end{thebibliography}
\bibliographystyle{IEEEtran}

\end{document}